\documentclass[aps,12pt]{revtex4}
\usepackage{epsfig}

\newcommand{\be}{\begin{equation}} 
\newcommand{\ee}{\end{equation}} 
\newcommand{\bea}{\begin{eqnarray}} 
\newcommand{\eea}{\end{eqnarray}} 
 
\newcommand\pubnumberk{LYCEN-2001-46\\ 
HIP-2001-35/TH\\}

\def\csumb{$^a$ IPN, Universit\'e de Lyon I, 
4 rue E.~Fermi, F-69622 Villeurbanne Cedex, France\\ 
$^b$ Physics Department, POB 64, FIN-00014, University of Helsinki, 
Finland} 
\newcommand\pubblock{\rightline{\begin{tabular}{l} 
                                                   \pubnumberk\end{tabular}}} 
\def\Title#1{\begin{center} {\Large\bf #1 } \end{center}} 
\def\Author#1{\begin{center}{ \sc #1} \end{center}} 
\def\Address#1{\begin{center}{ \it #1} \end{center}} 
 
\newenvironment{Abstract}{\begin{quotation}  }{\end{quotation}} 
 
\begin{document} 
\begin{titlepage} 
\pubblock 
\vfill 
\Title{$\bar{B}^0$ decays to $D^{(*)0} \eta$ and $D^{(*)0}\eta^\prime$} 
 
\Author{A. Deandrea$^{a}$ and A. D. Polosa$^{b}$} 
\Address{\csumb} 
\vfill 
 
\begin{Abstract} 
We consider the hadronic $B$ decays $\bar{B}^0\to D^{0} \eta$, 
$\bar{B}^0\to D^{*0} \eta$, $\bar{B}^0\to D^{0} \eta^\prime$, 
$\bar{B}^0\to D^{*0} \eta^\prime$ in the framework of a 
quark-flavour basis and factorization. The formalism allows to 
compute the decays to the $\eta$ meson and to relate them to those 
of the $\eta^\prime$. Measuring the branching ratios of these 
processes may shed light on the nature of the 
$\eta$--$\eta^\prime$ mixing. On the experimental side, only upper 
limits on the branching ratios are known at present. 
\end{Abstract} 
\vfill 
\end{titlepage} 
\eject \baselineskip=0.3in 
 
\section{Introduction} 
The hadronic $B$ decays $\bar{B}^0\to D^{0} \eta$, $\bar{B}^0\to 
D^{*0} \eta$, $\bar{B}^0\to D^{0} \eta^\prime$, $\bar{B}^0\to 
D^{*0} \eta^\prime$ are of considerable interest both 
theoretically and experimentally. Measuring the branching ratios 
of these processes may shed light on the nature of the 
$\eta$--$\eta^\prime$ mixing. On the experimental side, only limits on these 
decays are known at present from CLEO \cite{cleo}. A theoretical 
estimate of $\bar{B}^0\to D^{(*)0} \eta$, $\bar{B}^0\to D^{(*)0} 
\eta^\prime$ is important also because these processes are a 
background for the semi--inclusive processes 
$\bar{B}^0\to X_s \eta$, $\bar{B}^0\to X_s \eta^\prime$. For a recent 
experimental analysis, see \cite{babarnoi}. 
 
The processes $\bar{B}^0\to D^{(*) 0} \eta$, $\bar{B}^0\to D^{(*)0} 
\eta^\prime$ proceed via the internal spectator 
diagram shown in Fig.~1. They are 
colour suppressed since the colour of the quarks from the virtual 
$W$ has to match the colour of the produced $c$ quark from $b \to 
c$ and the spectator antiquark coming from the $B$ meson. Here we 
consider the quark component $\eta_q$ of the $\eta$ particle in 
the so called quark-flavour basis \cite{feldman}: 
\begin{equation} 
\left( \matrix {\eta\cr \eta^\prime \cr} \right)= 
\left( \matrix {\cos \phi & -\sin \phi\cr 
\sin \phi & \cos \phi \cr} \right) \; 
\left( \matrix {\eta_q \cr \eta_s \cr} \right) 
\end{equation} 
where the Fock state description of $\eta_{q,s}$ is: 
\begin{eqnarray} 
|\eta_q\rangle&=&\Psi_q \frac{|u\bar{u}+d\bar{d}\rangle}{\sqrt{2}}+...\\ 
|\eta_s\rangle&=&\Psi_s |s\bar{s}\rangle+... 
\end{eqnarray} 
The ellipses indicate higher Fock states, like glue states, while 
the wave functions $\Psi_{q,s}$ are the light-cone wave functions 
of the parton states. Our assumption is that the $\eta_q$ 
component is the one to be considered in the factorization diagram 
resulting from Fig. 1 (shrinking to a point the $W$ propagator). 
We will exploit the relation between $\eta$, $\eta^\prime$ and 
$\eta_q$ to compute the amplitude for the process $\bar{B}^0\to 
D^{(*)0}\eta^\prime$. 
 
We recall the form factor parameterization: 
\begin{eqnarray} 
\langle \eta(q_\eta)|V^{\mu}(q)|\bar{B}^0(p)\rangle &=& \left[ 
(p+q_\eta)^{\mu}+\frac{m_{\eta}^2-m_B^2}{q^2}q^\mu \right]\; 
F_1(q^2)\nonumber \\ &-& \left[ 
\frac{m_{\eta}^2-m_B^2}{q^2}q^{\mu} \right] \; F_0(q^2), 
\label{eq:effezero} 
\end{eqnarray} 
with $F_1(0)=F_0(0)$. Moreover we need the matrix elements: 
\begin{eqnarray} 
\langle {\rm VAC}|V^\mu| D(p) \rangle &=& if_D p^\mu \\\langle 
{\rm VAC }|V^\mu|D^*(\epsilon,p)\rangle &=& \epsilon^\mu m_{D^*} 
f_{D^*}, 
\end{eqnarray} 
and we will use the values \cite{neub} $f_D=200$ MeV, $f_{D^*}=230$ MeV. 
 
Adopting the Bauer-Stech-Wirbel approach to factorization 
\cite{bsw} we can consider the diagram in Fig. 1 describing the 
$\eta$ coupling through its $d\bar{d}$ component. The amplitude we 
need to compute is: 
\begin{eqnarray} 
G_{\bar{B}^0 D^0 \eta} &=& \langle \eta D^0 | H^{\rm 
BSW}|\bar{B}^0\rangle = \cos{\phi}\langle \eta_q D^0 | H^{\rm 
BSW}|\bar{B}^0\rangle\nonumber\\ &=&\frac{G_F}{\sqrt{2}} V_{cb} 
V_{ud}^* a_2 f_D  F_0^{(B\to\eta_q)}(m_D^2) (m_B^2-m_{\eta_q}^2) 
\frac{\cos{\phi}}{\sqrt{2}} \label{eq:penult}, 
\end{eqnarray} 
where $\phi$ is determined in \cite{feldman} to be 
$\phi=39.3^\circ$ and $a_2$ is the phenomenological coefficient of the 
Bauer-Stech-Wirbel effective Hamiltonian $a_2=0.29$. It is 
significantly larger than the leading--order result $a_2^{\mathrm LO} 
\sim 0.12$ corresponding to naive factorization. Such a large correction 
is not surprising. The QCD factorization formula of \cite{buch} 
can not be used to compute rigorously $a_2$ in these decays, however 
a rough estimate in the limit $m_c \ll m_b$ gives $a_2 \sim 0.25$ 
\cite{Neubert:2000ag}. The factor $1/\sqrt{2}$ in (\ref{eq:penult}) 
accounts for the fact that only the $d\bar{d}$ component enters in 
our diagram. The strange quark component is absent. 
 
The amplitude $G_{\bar{B}^0 D^0 \eta^\prime}$ is then connected to 
(\ref{eq:penult}) through: 
\begin{equation} 
G_{\bar{B}^0 D^0 \eta^\prime}=\tan{\phi} \; G_{\bar{B}^0 D^0 \eta}. 
\end{equation} 
In an analogous way we find: 
\begin{eqnarray} 
G_{\bar{B}^0 D^{*0} \eta}&=&\langle \eta D^{*0} | H^{\rm 
BSW}|\bar{B}^0\rangle= \cos{\phi} \langle \eta_q D^{*0} | H^{\rm 
BSW}|\bar{B}^0\rangle \nonumber\\ &=& \frac{G_F}{\sqrt{2}} V_{cb} 
V_{ud}^* a_2 f_{D^*} m_{D^*} F_1^{(B\to\eta_q)}(m_{D^*}^2) \; 
\epsilon\cdot(p+q_{\eta_q}) \;\frac{\cos{\phi}}{\sqrt{2}} \; . 
\label{eq:uno} 
\end{eqnarray} 
The term $\epsilon\cdot(p+q_\eta)$ is easily evaluated summing 
over the polarisations of $D^*$ once the square modulus of this 
amplitude is considered. 
 
\section{NS model} 
In order to get an idea about the amount of model dependence, we 
will consider two different models for the form factors. We start 
by considering a model by Neubert and Stech (NS) \cite{neub} which 
is a parameterization  based on simple assumptions, in reasonable 
agreement with the available data and with most theoretical 
predictions. In the case of heavy-to-light transitions, the form 
factors no longer obey the symmetry relations valid in the 
heavy-to-heavy decays. One can account for this by introducing, 
for each form factor, a function $\xi_i(w)$ replacing the 
Isgur-Wise function: 
\begin{eqnarray} 
F_1(q^2) &=& \frac{m_B + m_M}{2\sqrt{m_B m_M}}\;\xi_{1}(w) \,, \\ 
F_0(q^2) &=& \frac{2\sqrt{m_B m_M}}{m_B + m_M}\,\frac{w+1}{2}\;\xi_0(w) 
\,,  \end{eqnarray} 
where 
\begin{equation} 
w = v_B\cdot v_M = \frac{m_B^2 + m_M^2 - q^2}{2 m_B m_M} \, . 
\end{equation} 
The maximum value of $w=w_{\rm max}$ is obtained at $q^2=0$. 
For an estimate of the functions $\xi_i(w)$ a simple pole model 
is used: 
\begin{eqnarray} 
\xi_1(w)&=&\sqrt{\frac{2}{w+1}}\, \frac{1}{1+r} \,\frac{w_{\rm 
max}-w_1}{w-w_1} , \\ 
\xi_0(w)&=&\sqrt{\frac{2}{w+1}}\, \frac{1}{1+r\,\displaystyle 
\frac{w-1}{w_{\rm max}-1}} 
\end{eqnarray} 
where 
\begin{equation} 
r = \frac{(m_B-m_V)^2}{4 m_B m_V} \left( 1 
+ \frac{4 m_B m_V}{M_{0}^2 - (m_B - m_V)^2} \right) \,, 
\end{equation} 
where $V$ is the vector meson in the same doublet as the scalar meson 
involved in the decay described by the $F_0$ form factor (for the $\eta$ this 
corresponds to $\omega$). 
Moreover 
\begin{equation} 
w_1 = \frac{m_B^2+m_M^2-M_1^2}{2 m_B m_M} \,, 
\end{equation} 
where $M_i$ is the mass of the nearest resonance with the 
appropriate spin-parity quantum numbers (for $M_0=M_{B**}=5.754$ GeV it is the 
$0^+$ pole and for $M_1=M_{B*}=5.325$ GeV it is the $1^-$ pole). 
 
These form factors satisfy the relation 
\begin{equation} 
F_1(0) = F_0(0) \,, 
\end{equation} 
and scale as $\xi_i(w)\sim w^{-3/2}$ for large $w$, in accordance with the 
scaling rules obtained in \cite{Ali:1994vd}. 
 
The $B \to \eta$ form factors we obtain in this model are given in 
Table~1. The results for the branching ratios are given in Table~2. We also 
compute the corresponding $B \to \pi$ decays in order to allow a comparison 
with experimental data and theoretical predictions that is independent of the 
assumptions concerning the $\eta$ mixing. 
 
\section{CQM model} 
The computation of the form factor $F_0$ and $F_1$ can be carried 
out with the aid of a Constituent Quark Model (CQM) \cite{cqm}. 
This model has shown to be particularly suitable for the study of 
heavy meson decays. Since its Lagrangian describes Feynman rules 
for the vertices (heavy meson)-(heavy quark)-(light quark) 
\cite{rev}, transition amplitudes are computable via simple 
constituent quark loop diagrams in which mesons appear as external 
legs. We determine $F_0^{B\to\eta_8}$ considering the 
$\eta_8$ as a Goldstone boson. Therefore one has to take into 
account the relations between $\eta_8$ and $\eta_q$ for the 
process under study. To begin we observe that: 
\begin{equation} 
q^\mu \langle 
\eta_8|V^\mu|\bar{B}^0\rangle=(m_B^2-m_\eta^2)F_0^{(B\to \eta_8)}, 
\end{equation} 
where $q^\mu$ is the momentum carried by the current $V^\mu$. For 
our purposes $\eta$ is essentially the Goldstone octet component 
$\eta_8$: 
\begin{equation} 
\langle\eta|V^\mu|\bar{B}^0\rangle\simeq 
\langle\eta_8|V^\mu|\bar{B}^0\rangle. 
\end{equation} 
On the other hand the amplitude related to the process in Fig.1 
is: 
\begin{equation} 
\langle\eta|V^\mu|\bar{B}^0\rangle=\cos{\phi}\langle\eta_q|V^\mu|\bar{B}^0\rangle, 
\end{equation} 
if we also take into account a small mixing angle, $\theta_8$ in 
the notations of \cite{feldman}, we write: 
\begin{equation} 
\langle\eta_8|V^\mu|\bar{B}^0\rangle=\frac{\cos{\phi}}{\cos{\theta_8}} 
\langle\eta_q|V^\mu|\bar{B}^0\rangle. 
\end{equation} 
Again we stress that here we are neglecting the singlet component 
and we consider that the $B\to \eta$ matrix element which enters 
in the factorization diagram is the one containing only the 
$d\bar{d}$ components. In other terms we can write: 
\begin{equation} 
q^\mu \langle 
\eta_8|V^\mu|\bar{B}^0\rangle=\frac{\cos{\phi}}{\cos{\theta_8}} 
(m_B^2-m_{\eta_q}^2)F_0^{(B\to \eta_q)}, 
\end{equation} 
in such a way that: 
\begin{equation} 
F_0^{(B\to \eta_q)}=\frac{\cos{\theta_8}}{\cos{\phi}} 
\frac{(m_B^2-m_\eta^2)}{(m_B^2-m_{\eta_q}^2)}F_0^{(B\to \eta_8)}, 
\label{eq:ult} 
\end{equation} 
Notice that $\theta_8$ is determined in \cite{feldman} to be 
$\theta_8=-21.0^\circ$. For the $F_1$ form factor we obtain: 
\begin{equation} 
F_1^{(B\to \eta_q)}(m_{D^*}^2)=\frac{ 
\epsilon\cdot(p+q_{\eta_8})}{ 
\epsilon\cdot(p+q_{\eta_q})}\frac{\cos{\theta_8}}{\cos{\phi}} 
F_1^{(B\to \eta_8)}(m_{D^*}^2). \label{eq:due} 
\end{equation} 
 
We compute the $\eta$ particle 
via its $\eta_8$ component with a Goldstone like coupling in the 
meson-quark loop computation. The interaction Lagrangian 
generating this coupling has the form: 
\begin{equation} 
\bar{\psi}_a \gamma\cdot {\bf A}_{ab}\gamma_5 \psi_b, 
\end{equation} 
where the indices $a,b$ are the $u,d,s$ quark indices, $\psi$ 
being a triplet of flavour-$SU_3$. The structure of the ${\bf A}$ 
matrix is: 
\begin{equation} 
{\bf A}_\mu= -\frac{i}{2} (\xi\partial_\mu\xi^\dagger 
-\xi^\dagger\partial_\mu\xi), 
\end{equation} 
and $\xi=e^{i\pi/f}$. The ${\bf \pi}$ matrix has the well known form: 
\begin{equation} 
{\bf \pi}= \left(\begin{array}{ccc} 
\frac{\pi^0}{\sqrt{2}}+\frac{\eta_8}{\sqrt{6}} & \pi^+ &K^+\\ 
\pi^-&-\frac{\pi^0}{\sqrt{2}}+\frac{\eta_8}{\sqrt{6}} & K^0\\ K^- 
&\overline{K^0}& -\frac{2}{\sqrt{6}}\eta_8\end{array}\right) 
\end{equation} 
When the $d\bar{d}$ component is taken into account the Feynman 
rule for the vertex is $q^\mu \gamma_\mu\gamma_5/(\sqrt{6}f_\pi)$. 
We isolate three contributions to the $F_{0,1}$ form factors: a 
{\it non-derivative} contribution, a {\it direct} and a {\it 
polar} contribution \cite{noi}; see Fig. 2. The direct and the 
non-derivative contributions are represented in the CQM model by 
loop diagrams in which the internal lines are heavy quark and 
light quark propagators, while the external legs contain the 
incoming meson and the vector current. In the non-derivative 
diagram the $\eta$ external leg is attached at the same vertex 
(heavy quark)-(light quark)-(current) due to the structure of the 
CQM interaction Lagrangian. In the direct diagram the $\eta$ is 
instead attached to the light quark internal line (see 
\cite{noi}). The polar diagram allows for an intermediate polar 
state separating the current insertion from the loop diagram. Each 
of these diagrams contributes to the form factors $F_{0,1}$ in a 
calculable way. Summing up these contributions and imposing the 
condition $F_1(0)=F_0(0)$, which eliminates the spurious 
singularity in the form factor decomposition, one obtains: 
\begin{equation} 
F_i(q^2)= \alpha_j(q^2)(F_i^{\rm nd}+F_i^{\rm dir}(q^2)+F_i^{\rm 
pol}(q^2)),\label{eq:vedi} 
\end{equation} 
where $i=0,1$. The functions $\alpha_j(q^2)$, $j=0,1$ are needed 
to impose the condition $F_1(0)=F_0(0)$. This is because the sum 
in parenthesis reproduces the form factor on the lhs of the 
previous equation only at  $q^2_{\rm max}$, i.e., $\alpha(q^2_{\rm 
max})\simeq 1$. Following \cite{noi} we choose: 
\begin{equation} 
\alpha_j(q^2)=1+\alpha_j 
\frac{q^2-m_B^2-m_\eta^2}{2m_B\Lambda_\chi}=1-\alpha_j 
E_\eta/\Lambda_\chi, 
\end{equation} 
where $\Lambda_\chi=1$ GeV. In such a way we enforce that 
$\alpha(q^2_{\rm max})\simeq 1$. $E_\eta$ is the $\eta$ energy in 
the $B$ rest frame. We borrow $\alpha_0$ from \cite{noi} where we 
had $\alpha_0=0.27$. The form factor used in the computation of 
the branching ratios are of course those given in (\ref{eq:vedi}). 
The non-derivative (nd) contributions to the form factors are 
given by: 
\begin{equation} 
F_0^{\rm nd}= \frac{f_B}{\sqrt{6}f_\pi}~~~~~~~~~~~~ 
F_1^{\rm nd}=\frac{f_B}{2\sqrt{6} f_\pi}. 
\end{equation} 
The direct contributions are: 
\begin{eqnarray} 
F_{1}^{\rm Dir}(q^2)&=& \frac{S_1+S_2}{2}\\ F_{0}^{\rm 
Dir}(q^2)&=& \frac{S_1-S_2+A(q^2)(S_1+S_2)}{2A(q^2)}, 
\end{eqnarray} 
where: 
\begin{equation} 
A(q^2)=\frac{(m_B^2-m_\eta^2)}{q^2}, \nonumber 
\end{equation} 
and: 
\begin{eqnarray} 
S_1&=&\frac{2}{\sqrt{6}f_\pi}\sqrt{Z_H m_H}\left( \frac{m \; 
m_\eta^2}{m_H} Z+\left( \frac{2\; m}{m_H} \; v\cdot 
q_{\eta} -\frac{m_\eta^2}{m_H}\right)\Omega_1 -\frac{2 \; v\cdot 
q_{\eta}}{m_H}\Omega_4-\frac{2\; m_\eta}{m_H}\Omega_6\right)\nonumber\\ 
S_2&=&\frac{2}{\sqrt{6}f_\pi}\sqrt{Z_H m_H}(m^2 
Z-2\, m\, \Omega_1-m_\eta\, \Omega_2+2\, \Omega_3+\Omega_4-\Omega_5).\nonumber 
\end{eqnarray} 
In these expressions: 
\begin{equation} 
v\cdot q_\eta=\frac{m_B^2+m_\eta^2-q^2}{2 m_H}. 
\end{equation} 
The symbols $Z,Z_H,\Omega_i$ indicate integral expression listed 
in \cite{rev}. These are functions of $\Delta_H$, the difference 
between the heavy meson and the heavy quark constituent mass 
$m_H-m_Q$, $\Delta=\Delta_H-v\cdot q_\eta$ and $v\cdot q_\eta$, $v$ being 
the four velocity of the $B$ meson. $q^\mu$ is the momentum 
carried by the current insertion in the CQM loop diagram, $m$ is 
the constituent light quark mass, $m=300$ MeV. The 
suffix $H$ is referred to the $H=(0^-,1^-)$ multiplet predicted by 
heavy quark effective theory. 
 
The polar contributions are: 
\begin{eqnarray} 
F_{1}^{\rm pol}(q^2) &=& \frac{\hat{F}g}{\sqrt{6} f_\pi \sqrt{m_B}} 
\frac{1}{1-q^2/m_{B^*}^2}\\ 
F_{0}^{\rm pol}(q^2)&=&\frac{1}{m_B^2-m_\eta^2}\left( \frac{h 
m_\eta  \sqrt{m_B} {\hat F}^+}{\sqrt{6} f_\pi}\right) 
\frac{1}{1-q^2/m_{B^{**}}^2}, 
\label{eq:fzeropol} 
\end{eqnarray} 
where $B^{**}$ is the $0^+$ state of the multiplet $S=(0^+,1^+)$ 
while $B^*$ is the $1^-$ state of the $H$ multiplet. $h$ is the 
coupling constant $HS\pi$ and in CQM one finds $h=-0.76\pm0.13$ 
while $g=0.46\pm 0.04$, from $HH\pi$. We use the 
central values for all these constants and the value 
$\Delta_H=0.4$ GeV. Variations of $\Delta_H$ in the range 
$0.3-0.5$ GeV induce few $\%$ variations in the form factors (see 
Table 2). $\hat{F}$ and $\hat{F}^+$ are defined as the decay 
constants of the $H$ and $S$ heavy mesons respectively. 
 
Note that the form factors calculated in this section already 
include the factor taking into account that the process we are considering 
only involves the $d\bar{d}$ component of $\eta$ through the numerical 
coefficient of the Feynman rule. Therefore when combining (\ref{eq:penult}) 
and (\ref{eq:ult}) one has to be careful to remove a factor $1/\sqrt{2}$ 
from (\ref{eq:penult}): 
\begin{equation} 
G_{\bar{B}^0 D^0 \eta} =\frac{G_F}{\sqrt{2}} V_{cb} V_{ud}^* a_2 
f_D  F_0^{(B\to\eta_8)}(m_D^2) (m_B^2-m_{\eta}^2) 
\cos{\theta_8} 
\label{eq:prev} 
\end{equation} 
in terms of the $\eta_8$ state. 
{}From (\ref{eq:due}) and (\ref{eq:uno}) we get in a similar way: 
\begin{equation} 
G_{\bar{B}^0 D^{*0} \eta}=\frac{G_F}{\sqrt{2}} V_{cb} V_{ud}^* a_2 
f_{D^*} m_{D^*} F_1^{(B\to\eta_8)}(m_{D^*}^2) \epsilon\cdot(p+q_{\eta}) 
\cos{\theta_8} \; . 
\end{equation} 
Numerical results are given in Table~1 and 2. The two models give 
similar results for the form factors and branching ratios. The amount 
of model dependence due to the form factors is difficult to quantify. 
By comparing the two models one can conclude that they affect the 
branching ratios at the level of $\sim \pm 0.1 \times 10^{-4}$. The 
$a_2$ coefficient is considered fixed at the value 0.29 in the 
calculation. As it enters as $a_2^2$ in the branching ratio, by changing its 
value to 0.25 may add to the error on the branching ratio 
up to $\sim \pm 0.2 \times 10^{-4}$. However if the quark picture 
used in the calculation is correct such a large variation is not expected 
as $\bar{B}^0\to D^{0} \eta$ is in this case on the same footing as 
$\bar{B}^0\to D^{0} \pi$. 
 
\section{Conclusion} 
In order to test the idea of relating the $\eta$ and $\eta^\prime$ 
using the quark--flavour basis in the decays considered in this work, 
experimental data will have to measure the ratio of Br's $\eta^\prime 
/\eta$. If different from what estimated, it probably means 
that the quark mechanism (see Fig. 1) is not sufficient to explain 
the decays into $\eta^\prime$. This would suggest to look at some 
other mechanism, like the gluon anomaly explored in \cite{atw} 
where the large observed production of $\eta^\prime$ in $B\to 
X_s\eta^\prime$ decays has been studied. The mechanism 
suggested there is based on the subprocess $b\to s 
g^*\to s \eta^\prime g$ where the virtual gluon emerging from the 
standard model penguin couples to $\eta^\prime$ via a gluon 
anomaly vertex $g^*g\eta^\prime$. The possibility that the other 
gluon g is emitted by the light quark inside a $B$ meson has been 
examined in \cite{ahmadi}. The 
$B\to K \eta^\prime$ decay has been recently examined in the 
context of perturbative QCD in \cite{sanda}. In the case of $B\to 
D \eta^\prime$ studied here it is more difficult to imagine 
some gluonic mechanism of the kind described before. Anyway we 
cannot exclude this possibility since the $\eta^\prime$ can have a 
large glue component \cite{ahmadi}; see also \cite{norge}. 
Under the assumptions considered in this work $\bar{B}^0\to 
D^{0} \eta$  is similar to $\bar{B}^0\to D^{0} \pi$. The calculation is related 
to the one for $\bar{B}^0\to D^{0} \eta^\prime$ using the quark-flavour 
basis. We have checked that two different ways of computing the form factors 
give similar results. 
 
\section*{Note added in proofs}
Very recently there was a phenomenological determination of the $a_2$
coefficient~\cite{Cheng:2001sc} based on new data from Belle~\cite{newdata}.
This seems to be in contrast with all previous determinations based on the
factorization approach \cite{buch} addressing the problem of final--state
interaction in the $B \to D \pi$ channel. Such a problem could arise also in
this analysis where the $B\to D \eta$ is considered as the starting point, and
the factorization value for $a_2$ is assumed. We will consider this aspect in
future work.

\section*{Acknowledgements} 
We would like to thank R.~Gatto and A.~Hicheur for useful 
discussions. ADP acknowledges support from EU-TMR programme, 
contract CT98-0169. Institut de Physique Nucl\'eaire de Lyon 
(IPN Lyon) is UMR 5822.

\begin{table}[h] 
\begin{center} 
\caption{Values of the form factors used in the calculation. These 
are used in slightly different formulas for computing the 
amplitudes. The CQM form factors are indeed extracted using a 
Goldstone-like coupling for the $\eta$ particle and refer only 
to the $d{\bar{d}}$ component, while those of the NS model refer to the 
$u{\bar{u}} + d{\bar{d}}$ (see Sec.~3). In order to compare the two models on 
the same footing the form factors of the CQM model have to be multiplied by 
$\sqrt{2}$. 
} 
\label{tab:ff} 
\begin{tabular}{l l l l} 
\hline 
 &  $F_0^{B\to \eta} (0)$ & $F_0^{B\to \eta} (m_D^2)$ &  $F_1^{B\to \eta} 
(m_{D^*}^2)$ \\ 
\hline CQM Model & 0.15 & 0.17 & 0.23 \\ 
NS Model  & 0.26 & 0.28 & 0.33 \\ \hline 
\end{tabular} 
\end{center} 
\end{table} 
 
\begin{table}[ht] 
\begin{center} 
\caption{Theoretical predictions and 90\% C.L. upper limits on branching 
ratios. The errors quoted for the CQM model only refer to the variation 
of the parameters in the model and not to the model dependence of 
the result. 
}  \label{tab:predictions} 
\begin{tabular}{l l l l} 
\hline Decay Mode  &  @90\% C.L. \cite{cleo} & NS Model &  CQM 
Model \\ \hline $\bar{B}^0\to D^{0} \pi$ & $<$ 1.2$\times 10^{-4}$ 
& $0.77 \times 10^{-4}$ & $ 1.3^{+0.4}_{-0.3} \times 10^{-4}$ \\ 
$\bar{B}^0\to D^{*0} \pi$ & $<$ 4.4$\times 10^{-4}$ & $1.05 \times 
10^{-4}$ & $1.1 \pm 0.3 \times 10^{-4}$ \\ 
\hline 
$\bar{B}^0\to D^{0} \eta$ & $<$ 1.3$\times 10^{-4}$ & $0.50 \times 10^{-4}$ & 
$0.44 \pm 0.02\times 10^{-4}$ \\ 
$\bar{B}^0\to D^{*0} \eta$ & $<$ 2.6$\times 10^{-4}$ & $0.60 \times 10^{-4}$ & 
$0.70 \pm 0.04 \times 10^{-4}$ \\ 
$\bar{B}^0\to D^{0} \eta^\prime$ & $<$ 9.4$\times10^{-4}$ & $0.32 \times10^{-4}$ & 
$0.30 \pm 0.02 \times 10^{-4}$ \\ 
$\bar{B}^0\to D^{*0} \eta^\prime$ & $<$ 
14$\times10^{-4}$ & $0.41 \times10^{-4}$ & $0.47 \pm 0.04 \times 
10^{-4}$ \\ \hline 
\end{tabular} 
\end{center} 
\end{table} 
 
\begin{figure}[ht] 
\epsfysize=5truecm 
\begin{center} 
\epsffile{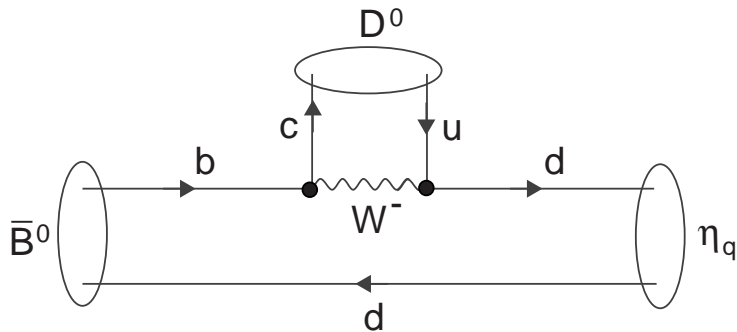} 
\caption{\label{fig:fig1} \footnotesize Flavour diagram for 
$\bar{B}^0\to \eta(\eta^\prime)$. The $\eta_q$ component allows to 
bridge between the two processes.} 
\end{center} 
\end{figure} 
 
\begin{figure}[ht] 
\epsfysize=6truecm 
\begin{center} 
\epsffile{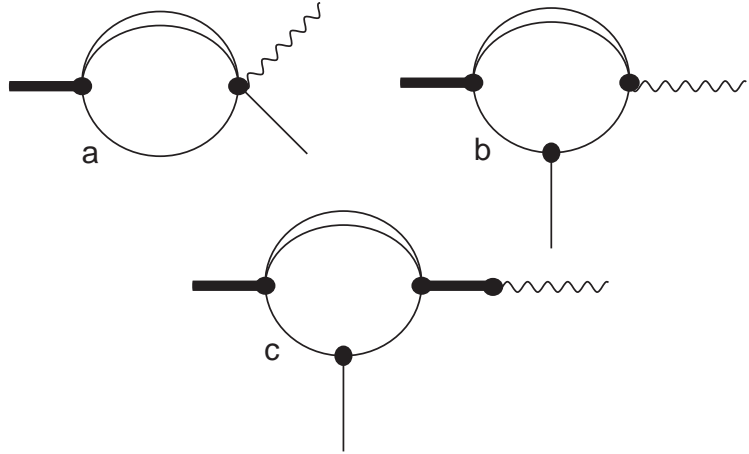} 
\caption{\label{fig:fig2} \footnotesize CQM diagrams. The heavy 
line represents the incoming heavy meson. The double line is the 
heavy quark, the curly line represents the current insertion. (a) 
is the {\it non-derivative} diagram in which the $\eta$ is coupled 
at the same vertex as the current, (b) is the {\it direct} diagram 
and (c) is the {\it polar} diagram: an intermediate heavy meson 
state is taken into account.} 
\end{center} 
\end{figure} 
 
\end{document}